\documentclass{jps-cp}
\usepackage{txfonts} 

\title{Multipole Modes for Triaxially Deformed Superfluid Nuclei}

\author{Kouhei \textsc{Washiyama}$^{1}$ and Takashi \textsc{Nakatsukasa}$^{1,2,3}$}

\inst{$^{1}$Center for Computational Sciences, University of Tsukuba, Tsukuba 305-8577, Japan \\
  $^{2}$Faculty of Pure and Applied Sciences, University of Tsukuba, Tsukuba 305-8577, Japan \\
$^{3}$iTHES, RIKEN, Wako 351-0198, Japan
}

\recdate{March 9, 2018}

\abst{To study shape fluctuations of nuclei in transitional regions,
the collective Hamiltonian method has often been employed.
We intend to construct the quadrupole collective Hamiltonian
with the collective inertial functions given by the local quasiparticle 
random-phase approximation (QRPA) based on the Skyrme energy density functional.
For this purpose, we first construct a practical framework of 
Skyrme QRPA for triaxial nuclear shapes with  
the finite amplitude method (FAM).
We show quadrupole strength functions for 
a triaxial superfluid nucleus $^{188}$Os
and the Thouless-Valatin
rotational moment of inertia by the local FAM-QRPA for $^{106}$Pd.
  }

\kword{shape fluctuation, collective Hamiltonian, energy density functional,
QRPA, finite amplitude method, triaxial nuclei, local QRPA, 
Thouless-Valatin moment of inertia}

\begin{document}
\maketitle

\section{Introduction}

Shape fluctuation in nuclei is an important aspect of quantum many-body physics
and is known to appear in transitional regions of the nuclear chart as, 
for example, shape coexistence and $\gamma$-soft nuclei\cite{heyde11}.
To investigate such shape fluctuation phenomena,
we need a theory to go beyond 
the mean field.
One of the promising methods is a five-dimensional quadrupole collective
Hamiltonian method.
Recently, the collective Hamiltonian method has been developed with 
modern energy density functionals (EDF)\cite{prochniak04,niksic09,delaroche10},
where the potential term is obtained by the constrained EDF
and the collective inertial functions
in the kinetic terms are estimated by the Inglis-Belyaev cranking formula
at each ($\beta,\gamma$) quadrupole deformation parameter.
Another progress has been made by deriving collective inertial functions 
by the local quasiparticle random-phase approximation (QRPA) 
including the dynamical residual interaction with
the pairing plus quadrupole (P+Q) force\cite{hinohara10, hinohara11,hinohara12}
based on the adiabatic selfconsistent collective coordinate 
method\cite{matsuo00, nakatsukasa12, nakatsukasa16}.
Our goal is to combine these two approaches, that is,
to derive collective inertial functions by the local QRPA 
with Skyrme EDF toward microscopic and non-empirical description of 
the collective Hamiltonian.
%
%
Solutions of the local QRPA require to construct a huge QRPA matrix
(its dimension of about $10^6$ for deformed nuclei)
and to diagonalize it
at each point of the ($\beta$,$\gamma$) plane.
This is computationally very demanding.

With the help of the finite amplitude method (FAM)\cite{nakatsukasa07, 
inakura09, avogadro11, stoitsov11,liang13,avogadro13,kortelainen15},
we have constructed a FAM-QRPA code 
for triaxial nuclear shapes.
Below, we will report quadrupole strength functions  
for a triaxially deformed superfluid nucleus $^{188}$Os.
Then, we will show applications of local FAM-QRPA to rotational moment
of inertia on the ($\beta,\gamma$) plane in a transitional nucleus $^{106}$Pd.

\section{Finite amplitude method}

The basic equation of FAM is the following linear-response equation
with an external field, 
\begin{subequations}\label{eq:FAM}
\begin{align}
  (E_\mu + E_{\nu} - \omega) X_{\mu\nu}(\omega) + \delta H^{20}_{\mu\nu}(\omega) &= -F^{20}_{\mu\nu}, \\
  (E_\mu + E_{\nu} + \omega) Y_{\mu\nu}(\omega) + \delta H^{02}_{\mu\nu}(\omega) &= -F^{02}_{\mu\nu},
\end{align}\end{subequations}
where $X_{\mu\nu}(\omega)$ and $Y_{\mu\nu}(\omega)$ are
FAM amplitudes at an external frequency $\omega$, 
$E_{\mu(\nu)}$ are one-quasiparticle energies, and $\delta H^{20,02}$
$(F^{20,02})$ are two-quasiparticle components of
an induced Hamiltonian (external field).
%
Details of the FAM formulation
can be found in Refs.\cite{nakatsukasa07, avogadro11}.
The important aspect of FAM is to use the linear response equation
and to replace a functional derivative
in the residual interaction with a finite difference form.
These avoid most time-consuming computation for QRPA, that is,
constructing QRPA matrix and diagonalizing it.
The solution of the FAM amplitudes in Eq.~(\ref{eq:FAM}) can be obtained by an
iterative procedure
(the modified Broyden method\cite{baran08} for our case).

We construct a FAM-QRPA code for iteratively solving the FAM equation (\ref{eq:FAM})
in three-dimensional (3D) Cartesian coordinate mesh.
More details of our 3D FAM-QRPA code can be found in Ref.\cite{washiyama17}.
Because of a specific reflection symmetry\cite{bonche85,bonche87} 
in the basis states,
the FAM equation is solved in only $x>0,y>0,z>0$ space.
We define the external isoscalar quadrupole operators with $K$ quantum numbers
as $Q^{(\pm)}_{2K}=(f_{2K}\pm f_{2-K})/\sqrt{2}$,
where $f_{2K} = (eZ/A)\sum^A_{i=1} r^2_i Y_{2K}(\hat{\boldsymbol{r}}_i)$
for $K>0$ and $Q^{(+)}_{20}=f_{20}$. 
In Ref.\cite{washiyama17}, we have confirmed the validity of 
our 3D FAM-QRPA code by comparing our results 
for axially symmetric nuclei with those in Refs.\cite{stoitsov11,kortelainen15}.

\section{Results}


\begin{figure}[tb]
\centering
\includegraphics[width=0.64\linewidth]{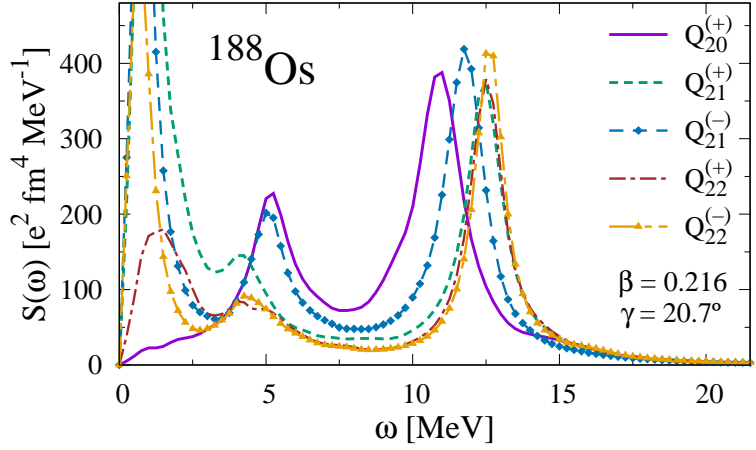}
\caption{Strength function $S(\omega)$ of different quadrupole modes
as a function of frequency $\omega$ for $^{188}$Os.
}
\label{fig:strength}
\end{figure}

We have reported the quadrupole strength functions of 
triaxially deformed superfluid nuclei, $^{110}$Ru and $^{190}$Pt, 
in Ref.\cite{washiyama17}.
Here, we show another example of triaxially deformed
superfluid nucleus, $^{188}$Os. The ground state is found
to be at $\beta=0.216,\gamma=20.7^\circ$ and superfluid in neutrons
when we use $19^3$ mesh, 1360 basis states, SkM$^*$ functionals, 
and volume pairing with the pairing strength 
that reproduces the neutron pairing gap of 1.25\,MeV in $^{120}$Sn.
We calculate the strength functions of the quadrupole modes as
\begin{align}
S(\omega) = -\frac{1}{\pi} \text{Im} \left(\sum_{\mu<\nu} F^{20*}_{\mu\nu} X_{\mu\nu}(\omega)+F^{02*}_{\mu\nu} Y_{\mu\nu}(\omega)\right),
\end{align}
obtained from converged FAM amplitudes
at 200 $\omega$ points between $\omega=0$ and 50\,MeV 
with $\Delta \omega =0.25$\,MeV.
The imaginary part of the frequency (of 0.5\,MeV)
is added as smearing width.

Figure \ref{fig:strength} shows the strength functions $S(\omega)$ of
quadrupole modes for different $K$ quantum numbers
as a function of frequency $\omega$ for $^{188}$Os.
We observe five $K$ splittings in strength and
three peaks for two $K=1$ and one $K=2$ associated with spurious modes 
of nuclear rotation around $x,y$, and $z$ axes near zero energy.
These are seen only in triaxial nuclei.
The energy-weighted sum rule 
of $K=0,2$ modes is well satisfied; 
98.6\% for both modes when $\omega$ is summed up to $\omega=50$\,MeV.

Next, we consider the evaluation of the collective 
inertial functions from the local QRPA with our 3D FAM-QRPA framework.
We start from the evaluation of the rotational moment of inertia.
Recently, the relation between the Thouless-Valatin inertia $M_{\textrm{NG}}$ of
a Nambu-Goldstone (NG) mode (spurious mode)
and the FAM strength function for the momentum operator 
$\hat{\mathcal{P}}_{\textrm{NG}}$ of this NG mode as an external field
at zero frequency was obtained in Ref.\cite{hinohara15}. This is expressed as 
\begin{align}\label{eq:NGmode}
  S(F=\hat{\mathcal{P}}_{\textrm{NG}}, \omega=0) &= \sum_{\mu<\nu}[F^{20*}_{\mu\nu} X_{\mu\nu}(\omega=0)+F^{02*}_{\mu\nu} Y_{\mu\nu}(\omega=0)] \notag \\
&=-M_{\textrm{NG}}\, .
\end{align}
%
%
For the case of nuclear rotation as an NG mode,
the corresponding momentum operator is the angular momentum operator,
and its Thouless-Valatin inertia is the rotational moment of inertia.
In Ref.\cite{petrik17},
this relation was used to obtain the Thouless-Valatin rotational 
moment of inertia with an axially symmetric FAM-QRPA. 
We use this relation for 
the Thouless-Valatin rotational moment of inertia. 
We perform the constrained HFB at each ($\beta,\gamma$) point,
and then perform FAM-QRPA calculation
on top of these constrained HFB states.  
This is the constrained HFB + local FAM-QRPA calculation.
In this case, we need to compute the FAM strength function at only $\omega=0$
without smearing width.
We take $^{106}$Pd as an example of such calculations.
To perform the constrained HFB + local FAM-QRPA at each ($\beta,\gamma$) point,
we use $17^3$ mesh, 1120 basis states, SkM$^*$ functionals, 
and volume pairing with the pairing strength 
that reproduces odd-even mass staggering in $^{106}$Pd.

\begin{figure}[tb]
\includegraphics[width=0.495\linewidth]{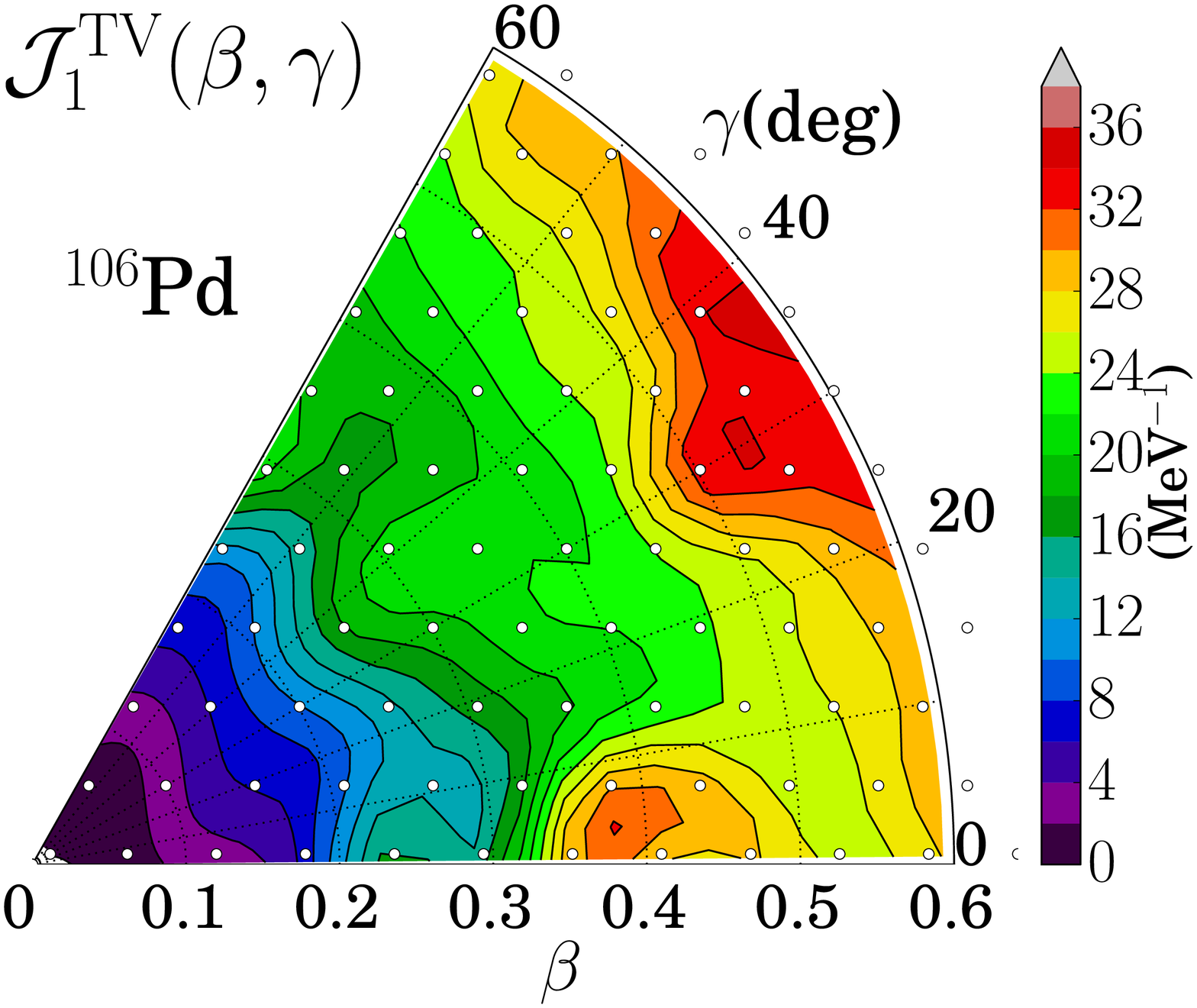}
\hspace{0.1em}
\includegraphics[width=0.495\linewidth]{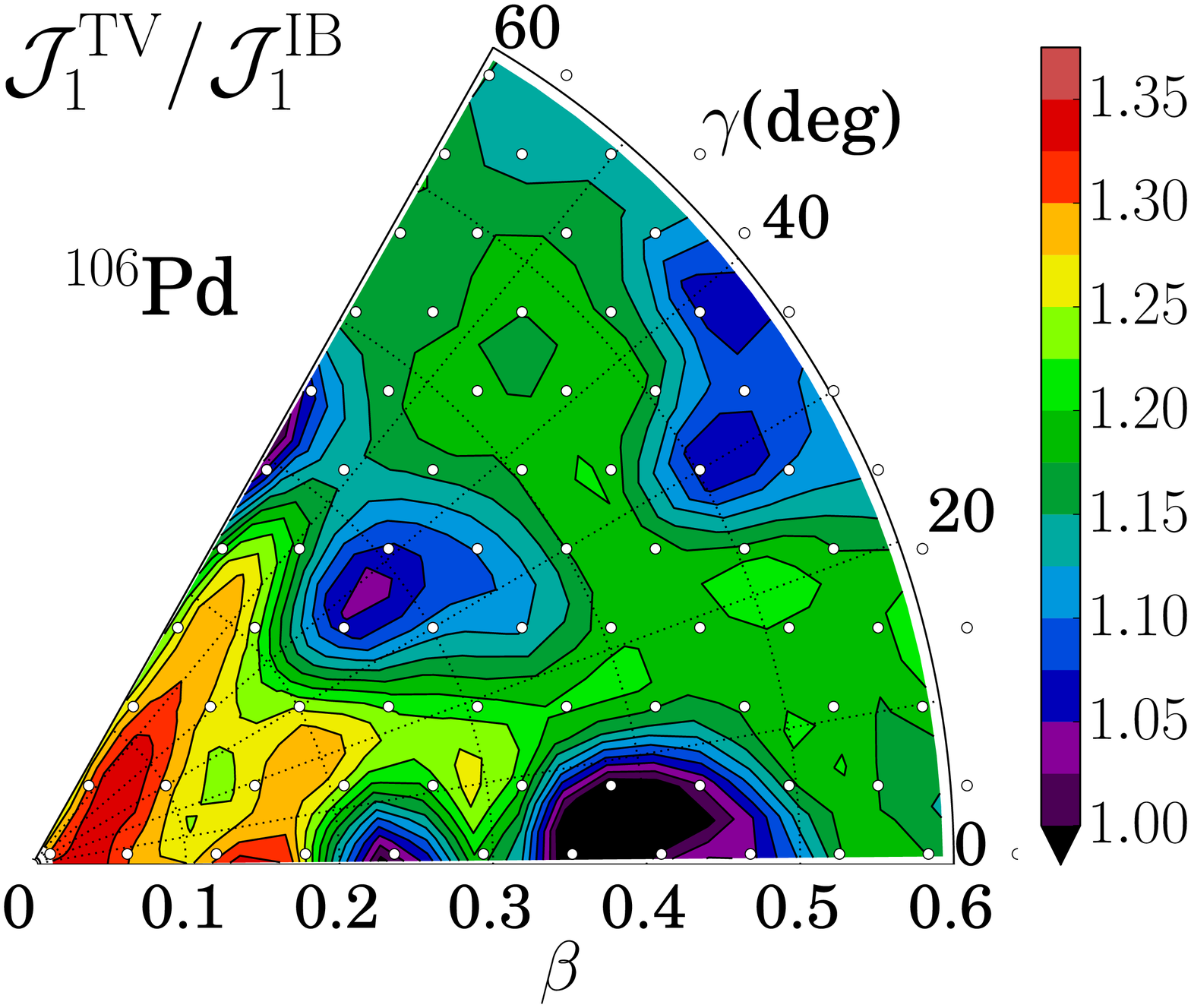}
\caption{(Left) Thouless-Valatin rotational moment of inertia on $x$ axis
$\mathcal{J}^{\textrm{TV}}_1(\beta,\gamma)$ 
as functions of $\beta$ and $\gamma$ for $^{106}$Pd.
(Right) Ratio of Thouless-Valatin rotational moment of inertia 
$\mathcal{J}^{\textrm{TV}}_1(\beta,\gamma)$
to Inglis-Belyaev one $\mathcal{J}^{\textrm{IB}}_1(\beta,\gamma)$
as functions of $\beta$ and $\gamma$ for $^{106}$Pd.
White circles on the ($\beta,\gamma$) plane represent the points at which
we perform constrained HFB + local FAM-QRPA. 
}\label{fig:moi}
\end{figure}

The left panel of Fig.~\ref{fig:moi} shows the
Thouless-Valatin rotational moment of inertia $\mathcal{J}_1(\beta,\gamma)$ 
on $x$ axis for $^{106}$Pd.
The value of the moment of inertia becomes larger with larger $\beta$,
as it should be. We find some structure such as local maximum of
moment of inertia at $\beta\sim 0.4,\gamma\sim 5^\circ$.
This is due to a change in a microscopic structure, especially 
in pairing correlations.
The right panel of Fig.~\ref{fig:moi} shows the ratio of the Thouless-Valatin
moment of inertia and the Inglis-Belyaev moment of inertia 
calculated by neglecting the residual interaction in FAM.
At most of the calculated ($\beta,\gamma$) points, this ratio exceeds unity.
This indicates the enhancement of the moment of inertia
by the residual interaction in FAM-QRPA calculations.
This is consistent with previous investigations\cite{hinohara10,hinohara12}.

\section{Summary}

Our goal is to construct the 5D quadrupole collective
Hamiltonian based on Skyrme EDF.
Toward this goal, we have first developed 3D Skyrme FAM-QRPA framework
for triaxial nuclear shapes.
We have obtained strength functions of quadrupole modes in triaxial superfluid
nucleus $^{188}$Os. These strengths show five $K$ splittings and 
three spurious peaks of nuclear rotations at zero energy as specific features 
of triaxial nuclei.
Then, we have extended our 3D Skyrme FAM-QRPA into the local one to estimate
the collective inertial functions for the 5D quadrupole collective Hamiltonian.
We have calculated the Thouless-Valatin moment of inertia in $^{106}$Pd
on the ($\beta,\gamma$) plane with the constrained HFB + local FAM-QRPA.
We have observed a significant enhancement of Thouless-Valatin 
moment of inertia from the Inglis-Belyaev one.

The important purpose of this work is to obtain FAM-QRPA results
with a small computational cost. 
We have obtained strength function at one $\omega$ point 
in Fig.~\ref{fig:strength}
within about 8 minutes in average 
and one value of moment of inertia at one ($\beta,\gamma$) point in
Fig.~\ref{fig:moi} within about 2 minutes,
with parallelization of 16 threads.

The estimation of the collective inertial functions in the vibrational 
kinetic term in the quadrupole collective Hamiltonian with the 3D FAM-QRPA is in progress. 
Then, the quadrupole collective Hamiltonian with the collective 
inertial functions by the constrained HFB + local FAM-QRPA
based on Skyrme EDF will be constructed in the near future.

\vspace{1em}
This work was funded by ImPACT Program of Council for Science,
Technology and Innovation (Cabinet Office, Government of Japan).
Numerical calculations were performed in part using the COMA (PACS-IX)
at the Center for Computational Sciences, University of Tsukuba, Japan.


\end{document}